\documentclass{Interspeech2024}

\usepackage[utf8]{inputenc}
\usepackage[T1]{fontenc}
\usepackage{adjustbox}
\usepackage{booktabs}

\interspeechcameraready

\title{Enhancing Crowdsourced Audio for Text-to-Speech Models}

\name[affiliation={1}]{Jos\'e}{Giraldo}
\name[affiliation={1}]{Mart\'i}{Llopart-Font}
\name[affiliation={1,2}]{Alex}{Peir\'o-Lilja}
\name[affiliation={1}]{Carme}{Armentano-Oller}
\name[affiliation={3}]{Gerard}{Sant}
\name[affiliation={1}]{Baybars}{Külebi}

\address{
  $^1$ Barcelona Supercomputing Center (BSC), Spain\\
  $^2$ Centre de Llenguatge i Computació (CLiC), Universitat de Barcelona, Spain
  $^3$ Universität Zürich, Switzerland
  }
  
\email{ \{jose.giraldo,marti.llopart,alexandre.peiro,carme.armentano,baybars.kulebi\}@bsc.es}


\begin{document}

\maketitle

\begin{abstract}  
High-quality audio data is a critical prerequisite for training robust text-to-speech models, which often limits the use of opportunistic or crowdsourced datasets. This paper presents an approach to overcome this limitation by implementing a denoising pipeline on the Catalan subset of Commonvoice, a crowdsourced corpus known for its inherent noise and variability. The pipeline incorporates an audio enhancement phase followed by a selective filtering strategy. We developed an automatic filtering mechanism leveraging Non-Intrusive Speech Quality Assessment (NISQA) models to identify and retain the highest quality samples post-enhancement. To evaluate the efficacy of this approach, we trained a state of the art diffusion-based TTS model on the processed dataset. The results show a significant improvement, with an increase of 0.4 in the UTMOS Score compared to the baseline dataset without enhancement. This methodology shows promise for expanding the utility of crowdsourced data in TTS applications, particularly for mid to low resource  languages like Catalan. 
\end{abstract}

\section{Introduction}

It is challenging to create multi-speaker text-to-speech (TTS) models because they rely on high-quality datasets reflecting the traits of each speaker. These types of datasets are both very scarce and not usually available for minority languages such as Catalan. Therefore, we had to find a way to recycle existing datasets for our task at hand. 

In recent years, crowdsourced datasets have seen a surge in quantity and quality, especially in Catalan \cite{armentano-oller-etal-2024-becoming-high}. More specifically, Commonvoice \cite{ardila2020commonvoicemassivelymultilingualspeech} stands out as a large-scale, crowdsourced and open-source collection of voice recordings created by the Mozilla Common Voice project. It is designed to provide a diverse and extensive range of voice data, and it's mainly used to train and improve automatic speech recognition (ASR) systems. Commonvoice audio files contain very rich information about each speaker's variability, a fundamental aspect needed to train multi-speaker TTS model. However, the samples from these datasets are noisy due to their recording conditions. In addition to background noise, recording artifacts and such, words are usually mispronounced \cite{naderi2023multidimensionalspeechqualityassessment}. Hence, the development of a strategy to take advantage of the existing crowdsourced data for TTS models is highly desirable in low to mid resource languages.

\subsection{Related works}

There have been some attempts to train multi-speaker TTS models with ASR datasets. For instance, Deep Voice 3 \cite{ping2018deepvoice3scaling} was trained with LibriSpeech (English ASR dataset) \cite{7178964}. The authors achieved this by using two preprocessing steps, involving firstly a standard denoising with SoX \cite{barras2012sox} and secondly a sentence segmentation step with Gentle \cite{Hawkins2024lowerquality}. In Cantonese, conditional diffusion models have also been used to enhance mel-spectrograms from an already existing ASR corpus. This technique includes text information and aims to address multiple types of audio degradation simultaneously \cite{tian2023diffusionbasedmelspectrogramenhancementpersonalized}. Other researchers trained the TF-GridNet speech enhancement model to low-resource datasets collected for ASR in Arabic, to then train a discrete unit-based TTS model on the enhanced speech \cite{ni2023exploringspeechenhancementlowresource}. 

Since it is not possible to manually evaluate and filter out the miriad of audio these datasets contain, there has recently emerged a viable solution to alleviate this problem: Deep learning (DL) models that leverage self-supervised representations to estimate the quality of the audio. These DL models have been tested to be superior than conventional metrics such as WBPESQ \cite{941023},  STOI \cite{5713237} and scale-invariant signal-to-distortion ratio (SI-SDR) \cite{roux2018sdrhalfbakeddone}, which provide objective evaluations. 

With the help of these novel automatic techniques, there are some papers where the authors instead opted for using crowdsourced datasets, successfully so. In previous research, a study used WM-MOS \cite{Andreev_2023} to filter the Commonvoice dataset with the objetive of training a multi-speaker TTS model for English \cite{10022766}. In their work, they used the pretrained DPTNet model of Asteroid \cite{pariente2020asteroidpytorchbasedaudiosource} for denoising. 
For the pan-African accented English, a study proposed to create a multi-speaker multi-accent TTS model with a crowdsourced dataset \cite{ogun20241000}. 
In a similar study, the Commonvoice dataset was used to create a multi-speaker TTS model for the Luganda language \cite{kagumire2024building}.
In this case, before training they denoised the dataset with a pre-trained speech enhancement model, \cite{defossez2020realtimespeechenhancement} and then filtered the audio files with WM-MOS.

Other studies trained multispeaker TTS models without removing the noise from the training datasets, and then applied a filter during the inference \cite{zhang2020denoispeechdenoisingtextspeech, 8683561}. Then, different groups opted for also encoding the characteristics of the environment within these models \cite{chang2022styleequalizationunsupervisedlearning}.
Other metrics, such as the word error rate (WER) or the Mel cepstral distortion, have been also used to automatically filter out noise from small datasets \cite{8639642,cooper17_interspeech}. Nonetheless, they haven't yet been applied to larger datasets. Unfortunately, apart from DL based models, recent attempts to automatically evaluate speech data have failed to perform well with previously unseen data \cite{Lo_2019,patton2016automoslearningnonintrusiveassessor}. This reason had extra weight with our project, where we are evaluating Catalan speech audio, a minority language.

 In comparison to all of these previous studies, we used Voicefixer \cite{Liu_2022}, which is designed to do audio restoration and has been proven to outperform single-task models at denoising and declipping at the time of starting this study. Then, we are using the Commonvoice Catalan dataset, which contains more hours and has more validation than any other language. This is a substantial reason for undertaking these denoising and filtering efforts, given that previous research succeeded with less data. Then, in our study, we used NISQA \cite{Mittag_2021} for automatic speech quality assessment. 

As a continuation, the second section of this paper relates the different open options for multi-speaker Catalan datasets. Section 3 then walks through the main components of the denoising pipeline while sections 4 explores how we executed this pipeline to train. Section 5 presents the results of our work and what consequences the different thresholds of NISQA have over the generated models.

\section{TTS datasets}

A TTS dataset is a collection of pairs of audio files and text transcriptions, some datasets include an additional column for the normalized text, however, the most important feature of a good TTS dataset is the quality of the audio files. These datasets don't contain background noises, have been recorded on a professional studio, and the prosody style is homogeneous per each speaker. Two notable TTS datasets for the Catalan language are Festcat \cite{bonafonte09_sltech} and OpenSLR69 \cite{openslr69}, containing respectively 22 and 5 hours of audio material. However, these corpora are substantially smaller in comparison to widely-used English TTS datasets, such as VCTK \cite{Yamagishi2019CSTRVC}, with 44 hours of speech and LJSpeech \cite{ljspeech17} with 24 hours.

Commonvoice \cite{ardila2020commonvoicemassivelymultilingualspeech} is a crowdsourced dataset which is mainly used to train ASR systems. Usually, the audio samples are polluted with background noises, the equipments used for the recording have low quality and a high variability between speakers. Commonvoice, in contrast with the previous Catalan TTS datasets contains more than 3000 hours of audio material.\\

\section{Methodology}

\subsection{Dataset}
For this paper we used the Commonvoice version 12, which contains 1,783,602 audio clips summing up to 2,721.97 hours of audio material. We use the validated set which contains 1,866 hours which we filtered to kept only the speakers that had more than 1400 seconds of audio, arriving finally to 826,900 clips.  It is important to note that 66\% of the corpus contains male samples, while female samples only amount for the remaining 33\%. Also, more than 40\% of the dataset contains samples from people older than 40 years, meaning that there's a low representation of the younger population in this dataset.

\subsection{Enhancement pipeline}

Due to the suitability to our dataset, we choose to use Voicefixer \cite{Liu_2022} for the enhancement phase. In an earlier exploration, we listened a random set of restored samples with Voicefixer, and the model was able to denoise and upsample low bandwidth audio from Commonvoice, A common problem with this dataset is the presence of click noises at the end of the recording which Voicefixer was able to remove.

\begin{figure}[h]
  \centering
  \includegraphics[scale=0.25]{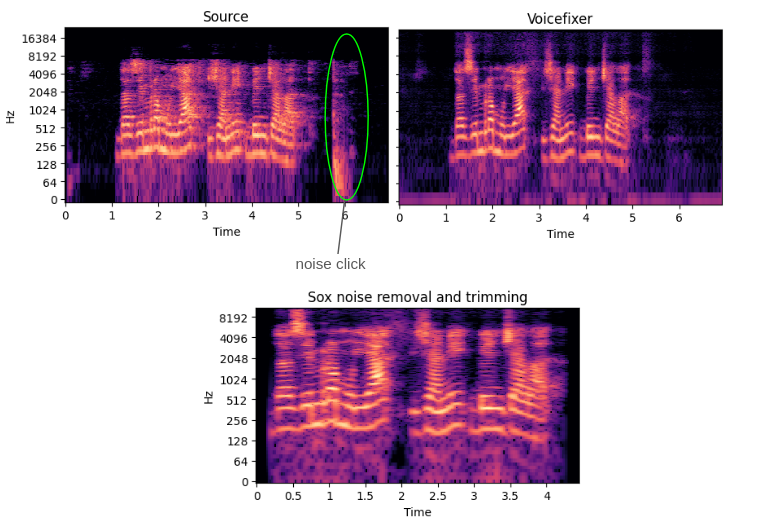}
  \caption{Spectrogram of an audio example, previously processed with the enhancement pipeline}
  \label{fig:enhancement}
\end{figure}

Despite the quality obtained with Voicefixer was quite good, there was a remaining noise as seen in The figure \ref{fig:enhancement} Because of that, we choose to remove it using spectral subtraction implementation from SoX \\

The final version of the audio enhancement pipeline has the following steps: 

\begin{enumerate}
    \item Format conversion: Mp3 to Wav using FFmpeg
    \item Enhancement model: Voicefixer enhancement model 
    \item Denoise: Get 0.5 seconds from the end of the audio. 
    Build a noise profile using the extracted 0.5 seconds.
    Perform Spectral Noise removal using the noise profile with Sox.
    \item Trimming:  Remove silences longer than 0.1 seconds and below -55dB from both the beginning and the end of the audio. Add padding of 0.1 seconds of silence at both the start and the end of the audio. \\

\end{enumerate}

The enhancement pipeline was optimized for CPUs, including the inference of the VoiceFixer model. For that, we implemented a parallel processing pipeline to maximize computational efficiency. This optimized configuration enabled us to process a substantial corpus of 826,900 audio samples in 48 hours. The system demonstrated a high-throughput performance, achieving an average processing rate of 4.5 audio files per second. This parallelization strategy significantly reduced the total processing time, making large-scale audio enhancement feasible on standard CPU infrastructures without the need for specialized GPU hardware.
\subsection{Enhancement model}

VoiceFixer is a unified framework for high-fidelity speech restoration. It can
restore speech from multiple distortions (e.g. noise, reverberation and clipping) and can expand degraded speech (e.g., noisy speech) with a low bandwidth, to 44.1 kHz full-bandwidth high-fidelity speech. This model uses a ResUNet to remove the distortions from  128 bin mel spectrograms, which are later used as inputs to a neural vocoder that output the final waveform.

\subsection{Filtering strategy}

We chose UTMOS \cite{saeki2022utmosutokyosarulabvoicemoschallenge} and NISQA \cite{Mittag_2021} to filter the dataset. NISQA was developed to asses the speech quality of an audio, and provides predictions for the quality dimensions of \textit{Noisiness}, \textit{Coloration}, \textit{Discontinuity}, and \textit{Loudness}. UTMOS, in contrast is a system trained to predict mean opinion score (MOS). Both models are open source\footnote{UTMOS: https://github.com/tarepan/SpeechMOS} \footnote{NISQA: https://github.com/gabrielmittag/NISQA} and available to use  \\

Using the aforementioned models, the quality scores were obtained per each audio, both in the enhanced and the original dataset. This allowed us to make a relationship between a certain quality threshold and the number of hours available. Although the NISQA Scores are distributed over all the range of values between 1 to 5, it's interesting to see that,in our enhanced dataset very few UTMOS values go higher than 3.0.

\begin{figure}[h]
  \centering
  \includegraphics[scale=0.35]{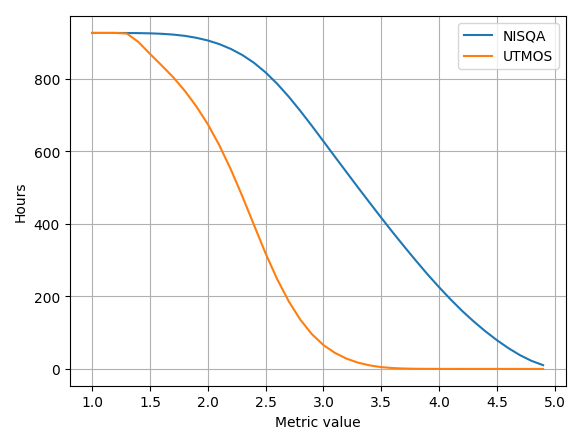}
  \caption{Relationship between hours in the enhanced dataset and different thresholds based on 2 quality metrics}
  \label{fig:utmos_nisqa}
\end{figure}

\section{Experimental setup}

With the filtering strategy, we built 4 subsets of 800, 200, 100 and 50 hours from the enhanced dataset, which respectively corresponds to NISQA values greater than 2.0, 4.0, 4.4 and 4.6. Rather than using the full NISQA scale, we opted to investigate the model's performance as we progressively reduced the dataset size, starting from the 200 hour subset (NISQA $\geq$ 4.0) as the breaking point for high quality audio. For that, we halved the duration in subsequent subsets to assess the impact of data quantity versus quality.\\
Additionally, we created a control subset by matching the audio samples from the 100 hour enhanced subset with their non-enhanced counterparts. This allows for a direct comparison between enhanced and non-enhanced data of equivalent content and duration.\\
These carefully curated subsets were then used to train separate TTS systems. Our objective was to evaluate the influence of data curation, based on automated audio quality metrics, on the final output quality of the generated speech samples. This experimental design enabled us to assess the trade-offs between data quantity, audio quality, and the effectiveness of our enhancement pipeline in improving TTS model performance.\\

We used Matcha-TTS \cite{mehta2024matchatts} as architecture for the TTS system, coupled with a custom espeak-ng \footnote{https://github.com/projecte-aina/espeak-ng}
backend, to convert input text to IPA phonemes. As a vocoder we used an adapted version of Vocos \cite{siuzdak2023vocos} that was pretrained with Catalan audio. We trained each of the experiments from scratch for 400K updates, using a single H100. The learning rate was set to 1e-4 for all the experiments, with a batch size of 32. Adamw optimizer was used without a scheduler.

\subsection{TTS model}

Matcha-TTS \cite{mehta2024matchatts} is an encoder-decoder architecture designed for fast acoustic modelling in TTS. The encoder part is based on a text encoder and a phoneme duration prediction, that together predict averaged acoustic features. The decoder essentially has a U-Net backbone inspired by Grad-TTS \cite{DBLP:journals/corr/abs-2105-06337}, which is based on the Transformer architecture. In the latter, by replacing 2D CNNs by 1D CNNs, a large reduction in memory consumption and fast synthesis is achieved.

\section{Results}

\subsection{Enhancement pipeline}

The enhancement phase resulted in a significant improvement in the Mean Opinion Score (MOS) dimension of the NISQA metric. We observed an increase of 0.3 points, raising the average mos score from 3.1 to 3.4.\\

\begin{table}[h]
\label{table:nisqa_bins}
\caption{NISQA Scores on enhanced and non-enhanced samples.}
\centering
\begin{tabular}{llll} 
\toprule
MOS scale & Original  & Enhanced & Diff \\
\bottomrule
1-2   & 1.78 & 1.79  & 0.01\\
2-3   & 2.53 & 2.64  & 0.11\\
3-4   & 3.45 & 3.48  & 0.03\\
4-5   & 4.39 & 4.40  & 0.01\\
\bottomrule
\end{tabular}
\end{table}

\begin{figure}[h]
  \centering
  \includegraphics[scale=0.14]{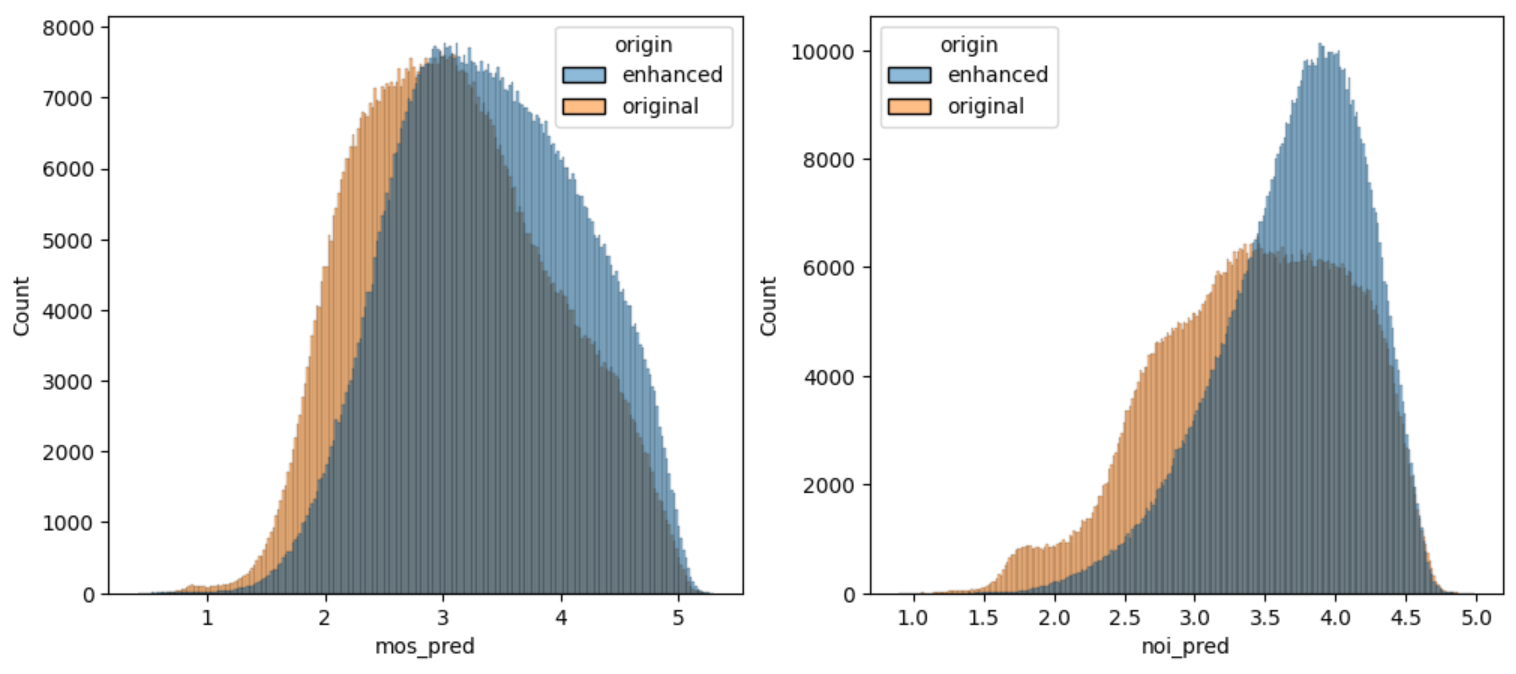}
  \caption{Distribution of NISQA scores over the datasets with and without enhancement.}
  \label{fig:nisqa_dist}
\end{figure}

The effect of the enhancement process is mostly noticeable in the lower-quality samples, as evidenced by the shift in the distribution of NISQA scores on Figure \ref{fig:nisqa_dist}. Specifically, the enhancement procedure induces a rightward shift in the lower end of the NISQA score distribution, increasing the scores of previously low-quality samples. Conversely, samples initially exhibiting high NISQA scores demonstrate minimal alteration, suggesting a ceiling effect or diminishing returns for already high-quality audio. This asymmetric impact on the score distribution underscores the targeted nature of the enhancement algorithm, preferentially improving low-quality samples while preserving the integrity of high-quality recordings.\\

A similar pattern is observed in the lowest quality samples, which may be attributed to the limitations of the enhancement algorithm in reconstructing severely degraded audio signals. This suggests that there exists a threshold of initial audio quality, below which the enhancement process becomes ineffective or potentially counterproductive.
\footnote{https://langtech-bsc.github.io/tts-enhancement-demo/}

\begin{table}[h]
\caption{NISQA Scores by sex variable.}
\centering
\begin{tabular}{llll} 
\toprule
Sex & Original  & Enhanced & Diff \\
\bottomrule
Female   & 3.19 & 3.379  & 0.185\\
Male   & 3.076 & \textbf{3.424}  & 0.348\\
\bottomrule
\end{tabular}
\label{table:nisqa_gender}
\end{table}

 As seen on Table \ref{table:nisqa_gender}  the enhancement technique appears more effective on male voices; despite starting lower, male audio achieves slightly better final quality.

\subsection{TTS Experiments}

We synthesized 15 phonetically balanced sentences using the top 20 speakers with the best quality from the dataset. We then employed the UTMOS model to get an score of each sentence, allowing us to evaluate the performance of each trained model. We also obtained the scores of PESQ, STOI and SI-SDR using the squim model, \cite{kumar2023torchaudiosquimreferencelessspeechquality}  which doesn't require a reference audio. \\ 

\begin{table}[h]
\caption{Comparison of metrics between TTS models trained with and without the enhancement pipeline}
\centering
\begin{tabular}{lllll} 
\toprule
Enhanced & UTMOS $\uparrow$         & PESQ   $\uparrow$       & STOI $\uparrow$ & SI-SDR    $\uparrow$      \\
\bottomrule
no   & $2.54_{\pm 0.06}$          & 3.19          & 0.95          & 20.14 \\
yes   & $\mathbf{2.94_{\pm 0.05}}$ & \textbf{3.38} & \textbf{0.98} & \textbf{22.61}  \\
\bottomrule
\end{tabular}
\label{table:enhanced_utmos}
\end{table}

As a first result, applying the enhancement pipeline on the dataset yields significant improvements across all objective audio quality metrics. Table \ref{table:enhanced_utmos} presents a comparative analysis of the TTS models trained with and without the enhancement pipeline. The enhanced dataset demonstrates superior performance across all measured parameters. Notably, UTMOS increased from 2.54 to 2.94, indicating a substantial improvement in perceived speech quality. The Perceptual Evaluation of Speech Quality (PESQ) score rose from 3.19 to 3.38, suggesting enhanced speech clarity and intelligibility. Short-Time Objective Intelligibility (STOI) improved from 0.95 to 0.98, reflecting increased speech intelligibility. Additionally, the Scale-Invariant Signal-to-Distortion Ratio (SI-SDR) showed a marked increase from 20.14 to 22.61, indicating a reduction in background noise. These quantitative improvements correlate with informal subjective listening assessments, further validating the efficacy of the enhancement pipeline.

\begin{table}[h]
\caption{Audio Quality metrics for different subsets of the dataset filtered with NISQA}
\centering
\begin{tabular}{lllll} 
\toprule
Hours & UTMOS $\uparrow$         & PESQ   $\uparrow$       & STOI $\uparrow$ & SI-SDR    $\uparrow$      \\
\bottomrule
800   & $2.82_{\pm 0.04}$          & 3.34          & 0.99 & 22.54           \\
200   & $2.92_{\pm 0.05}$          & 3.30          & 0.99 & 22.22           \\
100   & $\mathbf{2.94_{\pm 0.05}}$ & \textbf{3.38} & 0.98 & \textbf{22.61}  \\
50    & $2.90_{\pm 0.05}$          & 3.33          & 0.99 & 22.27           \\
\bottomrule
\end{tabular}
\label{table:duration_utmos}
\end{table}

Based on the data presented in Table \ref{table:duration_utmos}, we observe an improvement of the UTMOS score when the models are trained on the filtered high quality data(NISQA $\geq$ 4.0). Contrary to the intuitive assumption that larger datasets give superior results, our findings reveal that the 100-hour subset, curated through NISQA-based filtering, consistently outperforms other subsets across multiple quality metrics. This subset achieved the highest scores in UTMOS (2.94), PESQ (3.38), and SI-SDR (22.61). All subsets maintained high STOI scores (0.98-0.99), indicating robust intelligibility across different dataset sizes.\\

The observed decline in performance metrics for the 50-hour subset indicates a complex relationship between dataset size and audio quality filtering policy. For a TTS dataset with varying sample quality, there exists an optimal balance point between corpus size and filtering threshold. This optimum represents the most effective trade-off between data quantity and quality for maximizing TTS model performance. Our experiments suggest this optimal point in our dataset lies within the range of 50 to 200 hours of audio data, corresponding to NISQA thresholds between 4.0 and 4.6. This finding highlights the critical nature of precise dataset curation. Excessively strict filtering may lead to diminishing returns due to insufficient training data, while insufficiently rigorous filtering may introduce detrimental noise into the training process, compromising model quality.\\ 

\section{Conclusions}
We have presented an enhancement and data curation pipeline which is able to take advantage of crowdsourced datasets in order to train multispeaker TTS models in low to mid resource scenarios.
These results suggest that in TTS model training, the qualitative aspects of the dataset can be more critical than sheer quantity. Our NISQA-based filtering strategy demonstrates its effectiveness in distilling larger datasets into more refined, quality-centric subsets to reach improved audio output quality, potentially optimizing the trade-off between dataset size and model performance.

\section{Future Work}

Given the variability in performance obtained during the enhancement step, we believe the enhancement pipeline could be improved by making it adaptable to the specific distortions present in the audio. It would also be beneficial to test other enhancement models that can be scaled for large-scale inference. This approach could potentially lead to more consistent and effective results across a wider range of audio quality issues.

\section{Acknowledgements}

This work is funded by the Ministerio para la Transformación Digital y de la Función Pública and Plan de Recuperación, Transformación y Resiliencia - Funded by EU – NextGenerationEU within the framework of the project ILENIA with reference 2022/TL22/00215337. Development of the data and evaluation pipelines of this work has been promoted and financed by the Generalitat de Catalunya through the Aina project. Training was performed at Barcelona Supercomputing Center in MareNostrum 5.

\bibliographystyle{IEEEtran}
\bibliography{mybib}
\end{document}